\documentclass[aps,prl,twocolumn]{revtex4}  
\usepackage{graphicx}  
\usepackage{dcolumn}   
\usepackage{bm}            
\usepackage{amsmath}   
\usepackage{amssymb}   
\usepackage{hyperref}
\usepackage[latin1]{inputenc}
\usepackage{tikz}
\usetikzlibrary{shapes,arrows}
\usepackage{mathrsfs}

\newcommand{\eps}[0]{\varepsilon}
\newcommand{\pphi}[0]{\varphi}

\newcommand{\leqs}[0]{\leqslant}
\newcommand{\rarr}[0]{\rightarrow}
\newcommand{\rr}[0]{\mathbf{r}}

\newcommand{\footnoteremember}[2]{
\footnote{#2}
\newcounter{#1}
\setcounter{#1}{\value{footnote}}
}
\newcommand{\footnoterecall}[1]{
\footnotemark[\value{#1}]
}

\begin{document}

\title{How is the derivative discontinuity related to steps in the exact Kohn-Sham potential?}

\author{M.\ J.\ P.\ Hodgson}
\thanks{These authors contributed equally}
\affiliation{Max-Planck-Institut f\"ur Mikrostrukturphysik, Weinberg 2, D-06120 Halle, Germany}

\author{E.\ Kraisler}
\thanks{These authors contributed equally}
\affiliation{Max-Planck-Institut f\"ur Mikrostrukturphysik, Weinberg 2, D-06120 Halle, Germany}


\author{E.\ K.\ U.\  Gross} 
\affiliation{Max-Planck-Institut f\"ur Mikrostrukturphysik, Weinberg 2, D-06120 Halle, Germany}

\date{\today}

\begin{abstract}

The reliability of density-functional calculations hinges on accurately approximating the unknown exchange-correlation (xc) potential. Common (semi-)local xc approximations lack the jump experienced by the exact xc potential as the number of electrons infinitesimally surpasses an integer, and the spatial steps that form in the potential as a result of the change in the decay rate of the density. These features are important for an accurate prediction of the fundamental gap and the distribution of charge in complex systems.
Although well-known concepts, the exact relationship between them remained unclear. In this Letter, we establish the common fundamental origin of these two features of the exact xc potential via an analytical derivation. We support our result with an exact numerical solution of the many-electron Schr\"odinger equation for a single atom and a diatomic molecule in one dimension. Furthermore, we propose a way to extract the fundamental gap from the step structures in the potential.

\end{abstract}

\maketitle

Density functional theory (DFT)~\cite{HK64}, in the Kohn-Sham (KS) approach~\cite{KS65}, is widely used for simulating many-electron systems~\cite{PY,DG,Primer,EngelDreizler11,Burke12,Martin,Kaxiras03,Cramer04, ShollSteckel11}. The accuracy of density-functional calculations hinges on approximating the unknown exchange-correlation (xc) energy term, $E_\mathrm{xc}[n]$. While numerous successful approximations exist~\cite{PW92,Becke88,LYP,PBE96,PBEsol08, Sun_SCAN_PRL15,HSEsol,B3LYP1,B3LYP2,PerErnBurke96,ErnzerhofScuseria99, AdamoBarone99}, they often lack the discontinuous nature of the derivative of $E_\mathrm{xc}[n]$ with respect to electron number, $N$, at integer $N$ (derivative discontinuity (DD)~\cite{PPLB82,Cohen12,Baerends13,MoriS14,Mosquera14,Mosquera14a}).

The DD is essential for exactly describing the fundamental gap -- a feature of central importance for any material. When relying on the KS eigenvalues, approximate functionals (e.g.,~\cite{PW92,Becke88,LYP,PBE96}) lacking the DD underestimate this quantity by $\sim 50\%$~\cite{Tran07,Tran09,Eisenberg09,HSEsol,ChanCeder10,Lucero12}. Furthermore, these functionals may qualitatively fail in the dissociation limit, by predicting spurious fractional charges on the atoms of a stretched diatomic molecule
~\cite{PPLB82,Ossowski03,Dutoi06,Gritsenko06,MoriS06,Ruzsinszky06,Vydrov06,Perdew07,Vydrov07, KraislerKronik15}, 
violating the principle of integer preference~\cite{Perdew90}. 
This indicates that common approximations may also fail to describe charge transfer in molecules and materials~\cite{Perdew84,Tozer03,Maitra05,Toher05,Koentopp06,Ke07,Hofmann12,Nossa13,Fuks16}.

One manifestation of the DD is the emergence of a spatially uniform `jump', $\Delta$, in the level of the KS potential, $v_\mathrm{s}(\rr)$, as $N$ infinitesimally surpasses an integer value of $N_0$ by $\delta$: $\Delta = \lim_{\delta \rarr 0^+} v_\mathrm{s}(\rr;N_0 + \delta) - v_\mathrm{s}(\rr;N_0 - \delta)$~\cite{PerdewLevy83,ShamSchluter83,Kohn86,Godby87,Godby88,Chan99,AllenTozer02,Mundt05,Teale08, SagvoldenPerdew08,GoriGiorgiSavin08,Yang12,GouldToulouse14,Goerling15}. $\Delta$ originates from the piecewise-linearity of the total energy, $E(N)$~\cite{PPLB82,Cohen08,MoriS08,MoriS09,SteinKronikBaer_curvatures12,Kronik_JCTC_review12,Atalla16}, which implies a stair-step structure of the highest occupied (ho) KS eigenvalue, $\eps^\mathrm{ho}(N)$, with discontinuities at integer $N$. In particular, for a neutral system of $N_0$ electrons, infinitesimally below $N_0$, $\eps^\mathrm{ho}(N_0^-)$ equals $-I$, the negative of the ionization potential (IP), and infinitesimally above $N_0$, $\eps^\mathrm{ho}(N_0^+)$ equals $-A$, the negative of the electron affinity (EA)~\cite{PPLB82,LevyPerdewSahni84,PerdewLevy97,Harbola98,Harbola99,Yang12}. To enforce this behavior, the exact $v_\mathrm{s}(\rr)$ has to experience a discontinuous jump~\cite{PerdewLevy83,ShamSchluter83}:
\begin{equation}\label{eq:Delta}
\Delta = I - A - (\eps^\mathrm{lu} - \eps^\mathrm{ho}),
\end{equation}
where $\eps^\mathrm{ho}$ and $\eps^\mathrm{lu}$ are the ho and the lowest unoccupied (lu) KS eigenvalues at and infinitesimally below $N_0$ (from here on, the argument $N_0^-$ is suppressed for brevity). While this constant shift in $v_\mathrm{s}(\rr)$ does not affect the electron density, $n(\rr)$, it is vital to accurately predict the fundamental gap~\cite{PerdewLevy83,Perdew85,Bylander96,Stadele97,Stadele99,Magyar04,Rinke05,BeckeJohnson06,Gruning06, Gruning06a,Tran07,Rinke08,Tran09,Kuisma10,ChanCeder10,Zheng11,Kronik_JCTC_review12,ChaiChen13,ArmientoKummel13, Laref13,KraislerKronik13,KraislerKronik14,KraislerSchmidt15,Atalla16}.

In addition, the existence of the DD implies that the KS potential may form a `plateau' -- a constant increase in the level of $v_\mathrm{s}(\rr)$ in a given region -- to correctly distribute charge throughout the system~\cite{NATO85_Perdew_p302,Perdew90,KLI92, SagvoldenPerdew08, Karolewski09, Tempel09, Makmal11,Fuks11,Hofmann12,Nafziger15,GouldHellgren14,LiZhengYang15,KomsaStaroverov16,Kohut16}. At the edge of a plateau, $v_\mathrm{s}(\rr)$ forms a spatial step. 

In a stretched diatomic molecule $L \cdots R$, the height of the step, $S$, at the interface between the two atoms can be deduced~\cite{Hodgson16} from the change in the exponential decay of the density. When the density decay from the left atom, which has the analytic form $n(\rr) \propto e^{-2\sqrt{2I_L}|\rr|}$~\cite{PPLB82,PerdewLevy97}, meets the density decay from the right atom ($e^{-2\sqrt{2I_R}|\rr|}$), 
a step of height 
\begin{equation} \label{eq:S}
S = I_R - I_L + \eps^\mathrm{ho}_R - \eps^\mathrm{ho}_L
\end{equation}
forms in $v_\mathrm{s}(\rr)$ at this point~\cite{NATO85_AvB,RvL95,Gritsenko96,HelbigTokatlyRubio09, Hellgren12_PRA,YangZH14,Hodgson14,Hodgson16, BenitezProetto16}.

The exact relationship between the two aforementioned manifestations of the DD, $\Delta$ and $S$, remains unclear. These two quantities are usually treated as unrelated, because the jump in $v_\mathrm{s}(\rr)$ is a function of $N$, whereas a plateau is a function of space. Furthermore, the EA and the lu energy contribute to $\Delta$, while they are absent from $S$. 

A complete understanding of how the presence of steps in the KS potential account for the DD is essential for our ability to predict the fundamental gap of many-electron systems within DFT. 
Hence, in this Letter we directly address this problem by identifying the general mechanism that gives rise to $\Delta$ and $S$. We explore the properties of the plateau in the exact KS potential as a function of $N$ and show how $\Delta$ can be deduced from the step structure of the KS potential as $N$ decreases to an integer. This is done first via an analytical derivation, relying on fundamental properties of many-electron systems, and supported with an exact numerical solution of the many-electron Schr\"odinger equation in one dimension, for a single atom and a diatomic molecule. Modelling in 1D is necessary to solve the Schr\"odinger equation exactly; yet, the principles demonstrated may be generalized to 3D systems~\footnoteremember{ft:SupplMat}{See the Supplemental Material at [added by journal] for technical details.}.

To establish the relationship between $\Delta$ and $S$, we propose initially to study a single atom with a fractional $N = N_0 + \delta$. Then, the density is piecewise-linear~\cite{PPLB82}:
\begin{equation} \label{eq:n.piecewise}
 n(x;N) = (1-\delta) \cdot n(x;N_0) + \delta \cdot n(x;N_0+1),
\end{equation} 
being a combination of the density of a neutral atom and an anion ($N_0$ and $N_0+1$ electrons, respectively). When $\delta$ is small and positive, 
the density has \emph{two} regions of exponential decay:
as $|x| \rarr \infty$, $n(x;N) \propto n(x;N_0+1) \propto e^{-2\sqrt{2A}|x|}$ (the decay rate is governed by the IP of the anion, which equals the EA of the neutral). We term this the region of `$A$-decay'. However, for small $\delta$, $n(x;N_0) \propto e^{-2\sqrt{2I}|x|}$ starts to dominate the density at some point approaching the nucleus (`$I$-decay'). Based on Ref.~\cite{Hodgson16}, we expect a step in the KS potential to form at the crossover between these two decays. 

From the KS perspective, in the $A$-decay region $v_\mathrm{s}(x)$ reaches the asymptotic value $v'$, and in the $I$-decay region it has the value $v$. In general, $v$ and $v'$ may differ, forming the step $S:=v-v'$.

In the limit $\delta \rarr 0^+$, in the $A$-decay region, $n(x;N) \propto |\pphi^\mathrm{lu}(x)|^2 \propto e^{-2\sqrt{2(v'-\eps^\mathrm{lu})}|x|}$, where $\{ \pphi^i(x) \}$ are the KS orbitals. In the $I$-decay region, $n(x;N) \propto |\pphi^\mathrm{ho}(x)|^2 \propto e^{-2\sqrt{2(v-\eps^\mathrm{ho})}|x|}$. Therefore, $A = v'-\eps^\mathrm{lu}$ and $I = v-\eps^\mathrm{ho}$. Thus, $S=v-v'=I+\eps^\mathrm{ho}-(A+\eps^\mathrm{lu})=\Delta$, exactly as in Eq.~\eqref{eq:Delta}. 

To summarize, for an atom, when $N$ infinitesimally surpasses an integer, the plateau that forms in the region of the atom elevates the level of the KS potential by that required to obtain the exact fundamental gap.

We now model atoms in real space comprised of $N$ \textit{same-spin} electrons in 1D, where $1 \leqs N \leqs 2$, using the iDEA code~\cite{Hodgson13}~\footnoterecall{ft:SupplMat}. Solving the Schr\"odinger equation for $N \geqslant 2$ poses an immense computational challenge, which scales exponentially with $N$. Therefore, by enforcing the same spin for all electrons, we ensure that for a two-electron system \textit{two} KS orbitals are occupied (in contrast to a spin singlet -- one orbital occupied by two electrons with opposite spins), which is necessary for the concepts demonstrated in this Letter to be general~\footnote{These concepts do not rely on the electrons forming a singlet, but still apply in such a scenario.}. 
Our electrons interact via the appropriately softened, 1D Coulomb interaction $(\left | x-x' \right |+1)^{-1}$ (atomic  units)~\cite{Gordon05}\footnoterecall{ft:SupplMat}.
The external potential is $v_\mathrm{ext}(x) = -2/(0.4 \cdot \left | x \right |+1)$; it tends to zero as $|x| \rarr \infty$, as appropriate. We initially calculate $n(x)$ for $N=1$ and separately for $N=2$. Then, $n(x)$ for the $(1+\delta)$-electron system is calculated via Eq.~(\ref{eq:n.piecewise}).
Finally, we reverse engineer the exact KS potential from the exact ($1+\delta$)-electron density for varying $\delta$. 

\begin{figure}[htbp]
  \centering
  \includegraphics[width=0.925\linewidth]{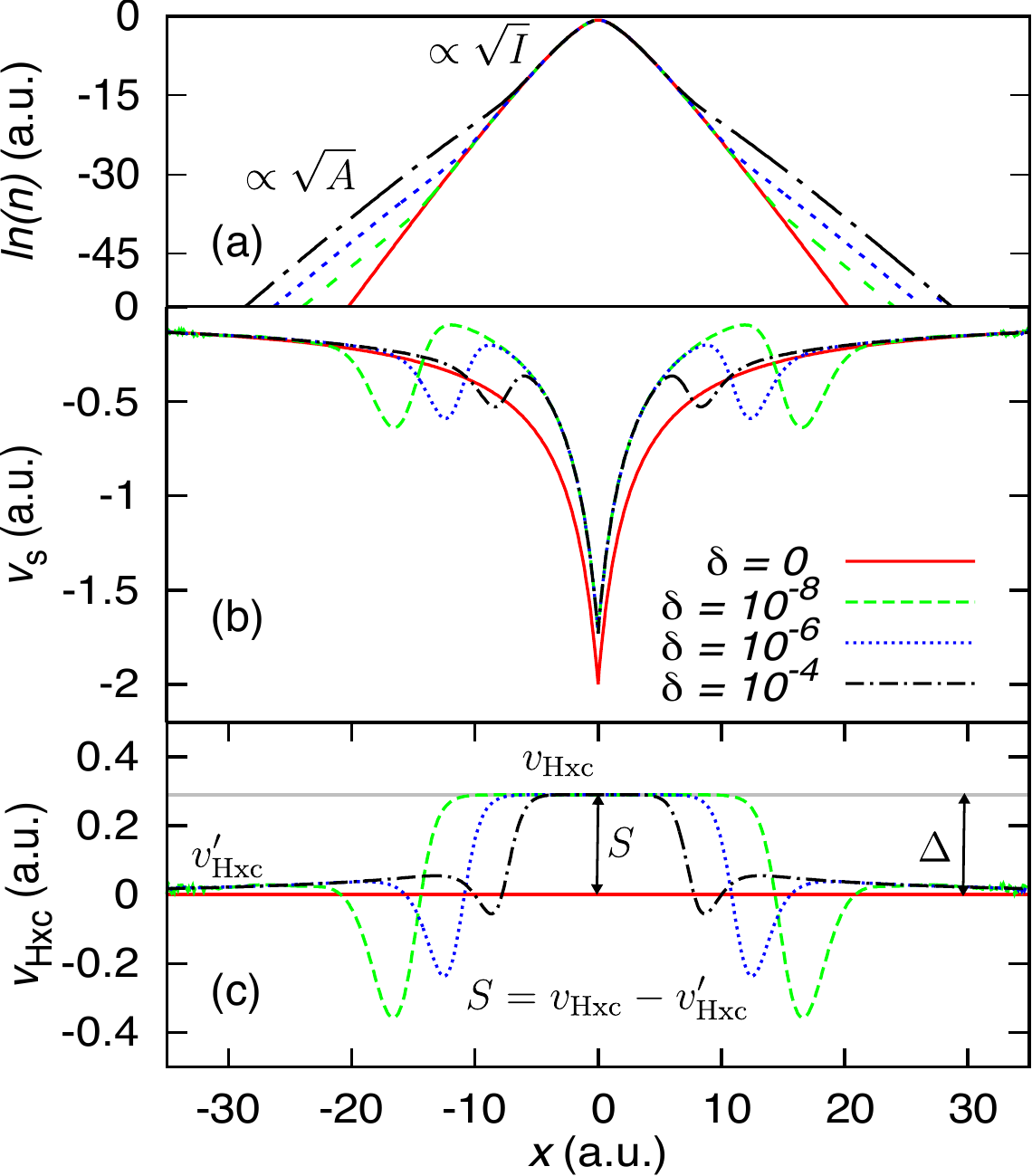}
\caption{
(a) $\ln{[n(x)]}$ for an atom with $1+\delta$ electrons; see key on (b). The $I$- and $A$-decay regions for $\delta>0$ are apparent. (b) Steps form in the exact KS potential, and (c) the plateau elevates the Hxc potential by $\Delta$, for $\delta>0$.}
\label{AtomL}
\end{figure}

Figure~\ref{AtomL}(a) shows the natural log of the density, $\ln{[n(x)]}$, for $\delta = 0, 10^{-8}, 10^{-6}$, and $10^{-4}$, as a function of $x$. Plotting the log helps to recognize the regions of exponential decay. For $\delta=0$, there is only one such region ($I$-decay). For $\delta > 0$, we clearly recognize the $I$- and the $A$-decay regions.
The above analysis indicates that the points where the decay rate changes are also the points where the steps in $v_\mathrm{s}(x)$ occur. This is demonstrated numerically in Fig.~\ref{AtomL}(b) and (c): for all finite values of $\delta$, the exact KS potential has two spatial steps. 
The steps act to elevate the level of the KS potential for the central region, where most of the electron density resides [cf.\ $v_\mathbf{s}(x=0)$ for $\delta=0$ and $\delta > 0$]. For all \textit{small} values of $\delta$ the step height is the same and equals $\Delta$, which we calculate from total energy differences and the exact KS eigenvalues; see the Hartree-exchange-correlation (Hxc) potential [$v_\mathrm{Hxc}(x)=v_\mathrm{s}(x) - v_\mathrm{ext}(x)$] in Fig.~\ref{AtomL}(c).
The positions of the steps vary with $\delta$ -- the smaller $\delta$ is, the further from the atom the steps form. Therefore, as $\delta \rarr 0^+$, the plateau becomes a spatially uniform shift in the potential, as required for the exact fundamental gap. 

The presented example clearly establishes the relationship between the derivative discontinuity and the steps' height in the atomic case, namely, that $\Delta = S$. It suggests that approximate xc functionals that are sensitive to the change in the decay rate of the density, and respond by forming steps in the xc potential (see suggestions in Refs.~\cite{ArmientoKummel13,Hodgson14,MoriS14}), are theoretically capable of producing the expected jump in the KS potential as $N$ surpasses an integer, and therefore yielding an accurate fundamental gap. 

We now consider a stretched diatomic molecule, which is \emph{one} system consisting of two atoms, $L$ and $R$, separated by a large distance, $d$. As $d \rarr \infty$, the number of electrons on each atom, $N_L$ and $N_R$, can be defined. We now imagine transferring an infinitesimal electronic charge, $\delta$, from $L$ to $R$. Therefore, $N_L = N^0_L - \delta$ and $N_R = N^0_R + \delta$, so the total number of electrons, $N_{L \cdots R} = N_L + N_R$ is constant. Relying on Eq.~\eqref{eq:Delta}, one may expect a plateau to form in the region of atom $R$, whose height is $\Delta_R=I_R-A_R - (\eps^\mathrm{lu}_R - \eps^\mathrm{ho}_R)$. Similarly, for a charge transfer from $R$ to $L$, one may expect a plateau $\Delta_L$ around $L$. However, Eq.~(\ref{eq:S}) demonstrates that the height of the step in $v_\mathrm{s}(\rr)$ that forms between $L$ and $R$ is independent of the EA of either atom. 

To consolidate these seemingly opposing viewpoints, we consider our atomic example above. We realize that for $\delta > 0$, we expect the density of atom $R$ to have two regions of exponential decay ($I_R$- and $A_R$-decay), while the density of atom $L$ will have one such region ($I_L$-decay only). For $\delta < 0$, the reverse picture applies.

\begin{figure}[htbp]
  \centering
  \includegraphics[width=1.0\linewidth]{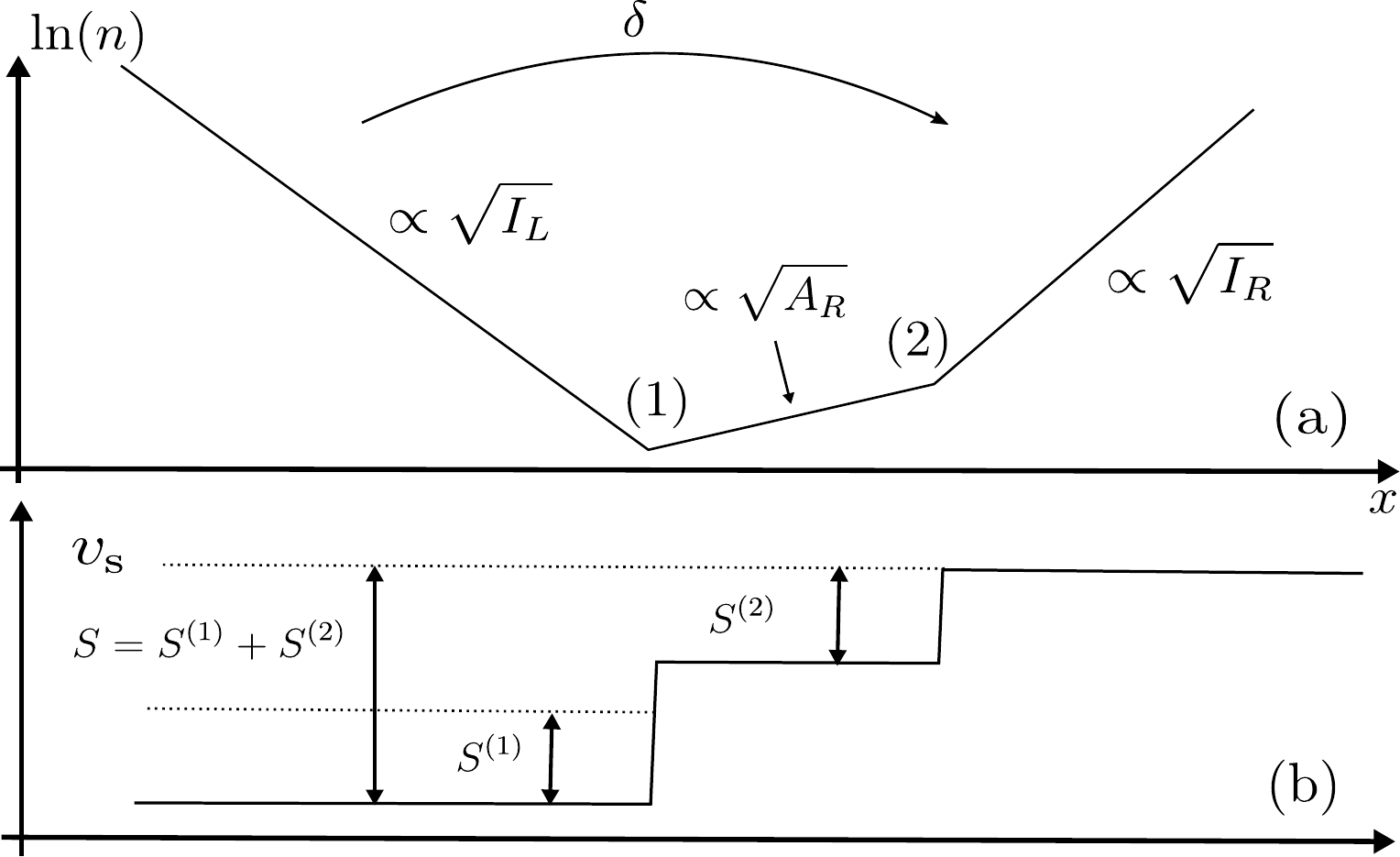}
\caption{(a) A diagram of $\ln{[n(x)]}$ far from, and between, the atoms of a molecule $L \cdots R$ shows  a transition from the $I_R$- to the $A_R$-decay region (point (2)) and from the $A_R$- to the $I_L$-decay region (point (1)). The changes in the density give rise to steps in the KS potential (b).}
\label{Decay_density}
\end{figure}

Figure~\ref{Decay_density}(a) shows a diagram of $\ln{[n(x)]}$ very far from, and between, the atoms. In this region the decaying density from atom $R$ changes its decay rate (because of the transferred charge) at point (2), then meets the decaying density from atom $L$ at point (1). Therefore, between the atoms there are two changes in the decay, and hence \emph{two steps} that manifest in the KS potential (Fig.~\ref{Decay_density}(b)). 
We emphasize that Step (1) arises because $L \cdots R$ is \emph{one} system, despite $d$ being large. Notably, the height of the step does \emph{not} depend on the magnitude of $n(x)$ at the point where it forms, and therefore the step is expected to appear at any $d$, as long as the atoms may be considered one system~\footnote{This situation is qualitatively different from the case of a dissociated molecule, originally introduced in Ref.~\cite{PPLB82} and recently discussed in Ref.~\cite{KraislerKronik15}, where one considers two \emph{completely} separated atoms with varying $N_L$ and $N_R$, at constant $N_{L \cdots R}$. Although $N_L$ and $N_R$ are varied in a concerted manner, strictly speaking this is no longer one molecule.}.
The heights of Steps (1) and (2) are derived as before: $S^{(1)}=A_R+\eps^\mathrm{lu}_R-(I_L+\eps^\mathrm{ho}_L)$ and $S^{(2)}=I_R+\eps^\mathrm{ho}_R-(A_R+\eps^\mathrm{lu}_R)$. They depend both on the IP, the EA, and the ho and lu KS eigenvalues. $S^{(2)}=\Delta_R$, whereas $S^{(1)}$ is not so recognizable a quantity. We term this the negative of the `charge-transfer derivative discontinuity', $\Delta_{L \rarr R}^\mathrm{CT}$, being the difference between the energy it costs to move an electron from $L$ to $R$ and the corresponding difference in the KS energy levels~\cite{Tozer03,Dreuw04,Hellgren12_PRA,Fuks13,Fuks14,Fuks16}. 

The overall difference in the level of $v_\mathrm{s}(x)$ in the region of $L$ and the region of $R$, which is the determining characteristic of $v_\mathrm{s}(x)$ regarding the distribution of charge between the atoms, is $S=S^{(1)}+S^{(2)}=I_R-I_L+\eps^\mathrm{ho}_R-\eps^\mathrm{ho}_L$, exactly as given by Eq.~(\ref{eq:S}), and independent of $A_R$ and $\eps^\mathrm{lu}_R$, in contrast to the individual gaps $S^{(1)}$ and $S^{(2)}$~\footnote{For $\delta < 0$, one obtains $S^{(2)}=-\Delta_L$, $S^{(1)} = \Delta_{R \rarr L}^\mathrm{CT}$, with the overall gap, $S$, remaining the same.}. 

\begin{figure}[htbp]
  \centering
  \includegraphics[width=1.0\linewidth]{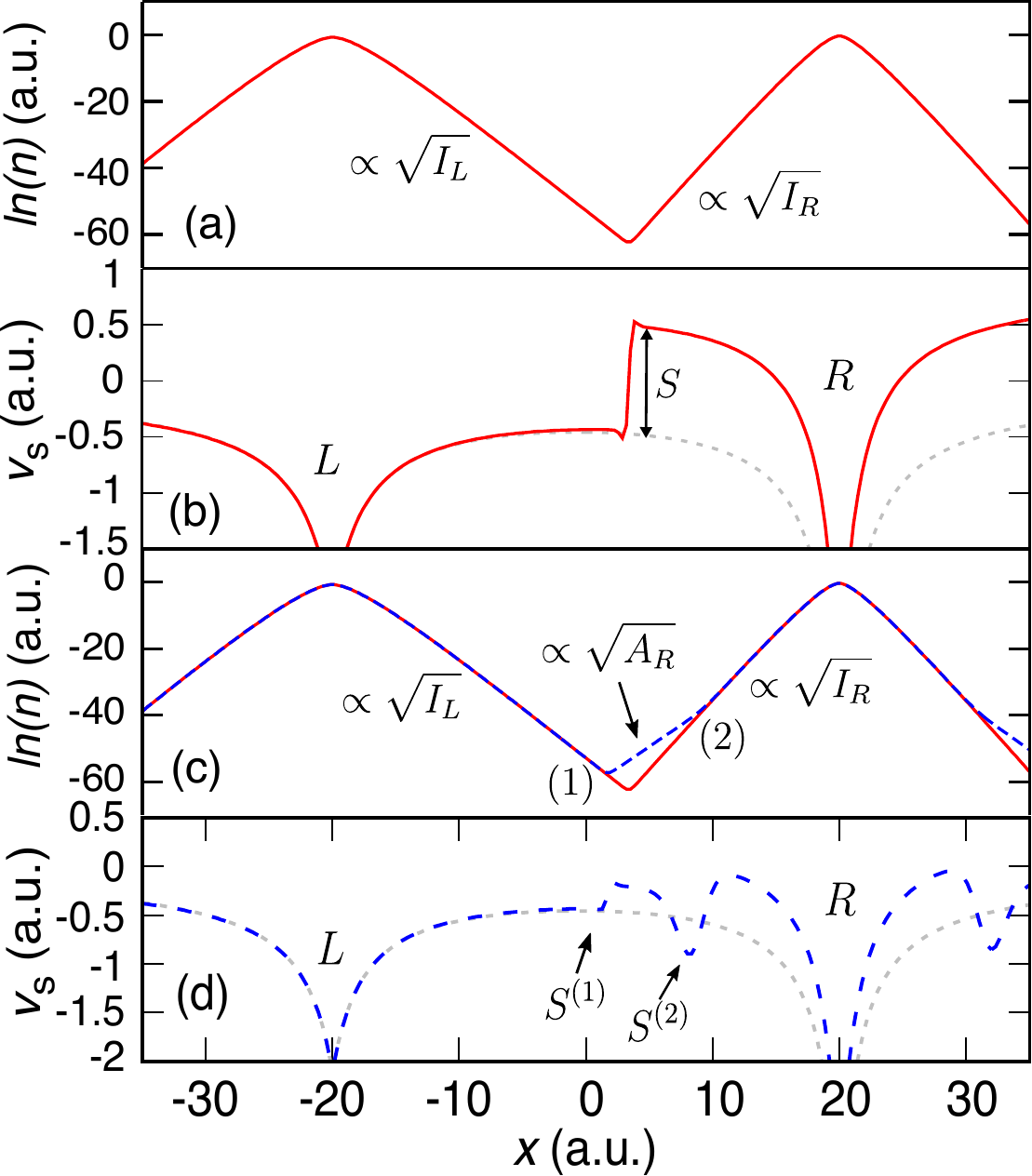}
\caption{(a) $\ln{[n(x)]}$ for 2 electrons in a stretched 1D molecule. (b) The corresponding exact KS potential (solid red) forms a step $S$ between atoms $L$ and $R$. The external potential (dotted gray) is shown for comparison. (c) $\ln{[n(x)]}$ for 2+$10^{-8}$ electrons (dashed blue) for the same external potential as (a). The 2-electron density of (a) is shown for comparison (solid red). (d) The exact KS potential corresponding to (c) (dashed blue) forms two steps between the atoms. The external potential (dotted gray) is shown for comparison.}
\label{Molecule}
\end{figure}

We now model a one-dimensional molecule $L \cdots R$ consisting of two atoms and two same-spin electrons ($N_{L \cdots R}=2$) separated by a large distance $d = 40$ a.u. 
The external potential, $v_\mathrm{ext}(x) = -4/(0.8 \cdot |x-\tfrac{1}{2}d|+1) -2/(0.4 \cdot |x+\tfrac{1}{2}d|+1)$, consists of two wells that represent the left and right nuclei. The potential is chosen so that  two \textit{interacting} electrons occupying this potential localize such that there is one electron's worth of charge in each well; see Fig.~\ref{Molecule}(a). To reproduce such a density in the KS system, $v_\mathrm{s}(x)$ must form a spatial step between the atoms~\cite{Hodgson16}. In the absence of the step, $\eps^\mathrm{lu}_R$ would be lower than $\eps^\mathrm{ho}_L$, which would cause both electrons to artificially localize on atom $R$.
Figure~\ref{Molecule}(b) shows a clear step in $v_\mathrm{s}(x)$, whose height is given by Eq.~(\ref{eq:S})
and its position is at the point where the decay rate of $n(x)$ changes [cf.\ Fig.~\ref{Molecule}(a) and~(b)]. To the far right of the atom $R$ there is another change in the decay rate of $n(x)$, which causes a step down (not shown).

For the above molecule we correctly observe no regions of $A$-decay, as none of the excited states of $L$ or $R$ are occupied. We now consider transferring a small amount of charge from $L$ to $R$ by making the right well deeper. This shifts the density minimum to the right, hence shifting the step in $v_\mathrm{s}(x)$ correspondingly, but no $A_R$-decay region forms. The height of the step is unaffected.

To further investigate the structure of $v_\mathrm{s}(x)$ when the density has \emph{three} regions of decay in between the nuclei, as shown in Fig.~\ref{Decay_density}, we increase $N_{L \cdots R}$ to be  $2 + \delta$, with $\delta \rarr 0^+$. We additionally calculate a three-electron system with the same $v_\mathrm{ext}(x)$ and use Eq.~(\ref{eq:n.piecewise}). In the three-electron case, one electron is localized on the left and two on the right. 
For $\delta = 10^{-8}$, the $A_R$-decay region can be observed in Fig.~\ref{Molecule}(c) and two points of decay change can be recognized in $\ln{[n(x)]}$ between the atoms, similar to Fig.~\ref{Decay_density}(a). 
Correspondingly, two steps in $v_\mathrm{s}(x)$ can now be observed between the atoms [Fig.~\ref{Molecule}(d)]. The sum $S^{(1)}+S^{(2)}$ is described by Eq.~\eqref{eq:S} and is independent of $A_R$. 

As $\delta$ decreases, points (1) and (2) travel towards each other until they meet, forming a density minimum. Likewise, steps (1) and (2) coincide and form the overall step $S$. In parallel, far from both atoms  the $A_R$-decay prevails (not shown). Hence, far to the left and to the right of the molecule we find a step $S_{L \cdots R} = I_L - A_R - (\eps_R^\mathrm{lu} - \eps_L^\mathrm{ho})$ that increases the level of the KS potential \emph{everywhere} for the molecule; we note $S_{L \cdots R} = -S^{(1)}$.
This is exactly the uniform jump predicted by Eq.~(\ref{eq:Delta}), noting that the global EA and lu level are those of atom $R$, whereas the global IP and ho level are those of atom $L$. Moreover, the DD of atom $R$, $\Delta_R$, can be deduced directly from the step structure of the KS potential at $\delta \rarr 0^+$, simply by adding $S$ and $S_{L \cdots R}$.

This example demonstrates that the sensitivity of approximate xc functionals to the change in the density decay rate and their ability to form steps in the xc potential is crucial to correctly distribute electronic charge in a system with appreciable spatial separation. Suggestions made, e.g., in Refs.~\cite{ArmientoKummel13,Hodgson14,MoriS14} are relevant in this context.

To conclude, in this Letter we clarified the relationship between the jump experienced by the Kohn-Sham (KS) potential as the number of electrons in the system infinitesimally surpasses an integer, and the spatial steps that form in the KS potential as a result of a change in the density decay rate. 

For a single atom, as the electron number passes an integer, two regions of exponential decay in the density manifest far from the nucleus. As a result, two steps in the exact KS potential form a plateau. In the limit of $\delta \rarr 0^+$, this plateau elevates the level of the potential everywhere in real space by $\Delta$, which is the amount required to obtain the exact fundamental gap from KS eigenvalues.

For a stretched diatomic molecule with an integer $N$, the KS potential forms a step between the atoms to correctly distribute the electronic charge in the molecule. The height of the step is independent of the atomic EAs. The step forms at the interface between the atoms, where the decay rate of the density changes. Slightly increasing the number of electrons above an integer causes the molecule's KS potential to develop two steps between the atoms. Each step individually depends both on the atomic IP and EA; however, the overall difference in the level of the KS potential between the atoms remains insensitive to the atomic EAs.   

The properties of the exact KS potential outlined in this work depend on the fine details of the asymptotic decay of the electron density. Accounting for these properties with standard approximations to the exchange-correlation (xc) functional poses a great challenge. However, since an accurate step structure in an approximate xc potential is crucial to predict the fundamental gap and provide a correct distribution of the electronic charge in complex systems, it should be taken into account in the development of future approximations.

\acknowledgments{We acknowledge Rex Godby for providing us with computational resources. E.K.\ greatly appreciates the support of the Alexander von Humboldt Foundation.}  

\bibliography{bibliography}

\end{document}


\title{How is the derivative discontinuity related to steps in the exact Kohn-Sham potential?\\
Supplementary Material}

\date{\today}

\author{M.\ J.\ P.\ Hodgson}
\thanks{These authors contributed equally}

\author{E.\ Kraisler}
\thanks{These authors contributed equally}

\author{E.\ K.\ U.\  Gross} 
\affiliation{Max-Planck-Institut f\"ur Mikrostrukturphysik, Weinberg 2, D-06120 Halle, Germany}

\maketitle

\section{The iDEA code}

The electron densities presented in the Letter were found by solving the many-electron Schr\"odinger equation in one-dimension with the iDEA code \cite{PhysRevB.88.241102}. The code finds the fully-correlated, ground-state, many-electron wavefunction by propagating an exchange-antisymmetric wavefunction through imaginary time in our chosen external potential \cite{doi:10.1080/01621459.1949.10483310}. The Crank-Nicolson method \cite{crank-nicol} is used to numerically propagate the wavefunction.
The interaction between the electrons is not the full Coulomb interaction, $|\mathbf{r}-\mathbf{r}'|^{-1}$, but rather the softened Coulomb interaction, whose form has been stated in the main text. The softened Coulomb interaction is appropriate for one dimension, and avoids the numerical instabilities that arise in 1D. This softened Coulomb interaction has been successfully used in the past \cite{PhysRevA.72.063411,PhysRevB.88.241102}, and proved useful in modelling atoms and molecules in one dimension,
in various contexts.
The numerical conversion of the electron density is monitored by evaluating the integral $C_n = \int_{-\infty}^{\infty} \left | n(x)-n'(x) \right | \mathrm{d}x$. The density is considered converged once $C_n < 10^{-14}$ a.u.
Our solution of the many-electron Schr\"odinger equation also gives access to the exact energy of the system. The energies are converged to three significant figures.

From the exact density we can obtain the exact Kohn-Sham (KS) potential ($v_\mathrm{s}$) by inversion. For this we use the `reverse-engineering' algorithm of iDEA \cite{PhysRevB.88.241102}, where $v_\mathrm{s}\rightarrow v_\mathrm{s} + \mu[n(x)^p-n'(x)^p]$ is iterated over until $[n(x)^p-n'(x)^p]$ is minimzed. $p$ and $\mu$ are numerical parameters; $p$ is typically $0.05$ and focuses the iterative procedure on the low density regions (that we are interested in), and $\mu$ is used for numerical stability (typically $\mu=1$). 
From the KS potential we can obtain the KS eigenvalues and orbitals by solving the one-electron Schr\"odinger equation.

We use same-spin electrons in our calculations meaning they each occupy a distinct KS orbital. In this way, we maximize the number of KS orbitals occupied in the system for a given computational effort, hence allowing higher energy states to be occupied and making our results more general.

\subsection{Atomic systems}

Converged results were obtained for both the one- and two-electron atomic systems with spatial grid spacing of $\delta x = 0.4$, however, further converged results were presented in the Letter ($\delta x = 0.1$). Our results for $N=1+\delta$, where $\delta = 10^{-8}, 10^{-6}$, and $10^{-4}$, are converged also, as they are a piecewise-linear combination of the two integer densities. 

The atomic systems share the external potential given in the main text. For both the one- and two-electron systems, the electrons are bound in the well, and have a well-defined ionization potential ($I$) and electron affinity ($A$). 

The exact KS potential is reverse-engineered using the exact electron density (described above). The absolute integrated difference between the KS density, $n'(x)$, and the exact, $n(x)$, $C_n$ above, is below $10^{-12}$ a.u.\ for all our atomic systems, hence the potential that yields $n'(x)$ is deemed exact. 

\subsection{Diatomic molecule systems}

Converged results were obtained for the one-, two-, and three-electron diatomic molecular systems with $\delta x=0.4$; furthermore, the results presented in the Letter correspond to $\delta x=0.32$. Hence, as for the atomic systems, the $N=2+10^{-8}$ diatomic molecule is also converged. Each atom in the diatomic molecule has a well-defined `local' ionization potential and electron affinity, and $I$ and $A$ for the whole system are also well defined. The atoms are stretched sufficiently ($d=40$ a.u.) to ensure that the electron density from each atom decays asymptotically. 

The exact KS potential for both diatomic molecular systems yields $C_n<10^{-12}$ a.u. 

\bibliography{Bibtex}